\documentclass[twocolumn,showpacs,preprintnumbers,amsmath,amssymb, superscriptaddress, prb]{revtex4}
\usepackage{graphicx}
\usepackage{dcolumn}
\usepackage{bm}
\usepackage{textcomp}

\usepackage[hyperfigures=true,bookmarksnumbered=true,bookmarksopen=true,bookmarks=true,
linkcolor=blue,urlcolor=blue,citecolor=blue,colorlinks=true,pdfhighlight=/O,pdfstartview={XYZ null null 1}]{hyperref}

\begin{document}


\title{Magneto-elastic couplings in the distorted diamond-chain compound azurite}


\author{Pham Thanh Cong}
\affiliation{Physics Institute, Goethe University Frankfurt(M),
D-60438 Frankfurt(M), Germany}

\author{Bernd Wolf}
\email[]{wolf@physik.uni-frankfurt.de} \affiliation{Physics
Institute, Goethe University Frankfurt(M), D-60438 Frankfurt(M),
Germany}

\author{Rudra Sekhar Manna}
\affiliation{Physics Institute, Goethe University Frankfurt(M), D-60438 Frankfurt(M), Germany}

\author{Ulrich Tutsch}
\affiliation{Physics Institute, Goethe University Frankfurt(M), D-60438 Frankfurt(M), Germany}

\author{Mariano de Souza}
\affiliation{Physics Institute, Goethe University Frankfurt(M), D-60438 Frankfurt(M), Germany}\affiliation{Instituto de Geoci\^encias e Ci\^encias Exatas - IGCE, Unesp - Univ  Estadual Paulista, Departamento de
F\'isica, Cx.\,Postal 178, 13506-970 Rio Claro (SP), Brazil}

\author{Andreas Br\"{u}hl}
\affiliation{Physics Institute, Goethe University Frankfurt(M), D-60438 Frankfurt(M), Germany}

\author{Michael Lang}
\affiliation{Physics Institute, Goethe University Frankfurt(M), D-60438 Frankfurt(M), Germany}


\begin{abstract}
We present results of ultrasonic measurements on a single crystal of
the distorted diamond-chain compound azurite
Cu$_3$(CO$_3$)$_2$(OH)$_2$. Pronounced elastic anomalies are
observed in the temperature dependence of the longitudinal elastic
mode $c_{22}$ which can be assigned to the relevant magnetic
interactions in the system and their couplings to the lattice
degrees of freedom. From a quantitative analysis of the magnetic
contribution to $c_{22}$ the magneto-elastic coupling $G$ =
$\partial J_2$/$\partial \epsilon_b$ can be determined, where $J_2$
is the intra-dimer coupling constant and $\epsilon_b$ the strain
along the intra-chain $b$ axis. We find an exceptionally large
coupling constant of $|G| \sim ($3650 $\pm$ 150)\,K highlighting an
extraordinarily strong sensitivity of $J_2$ against changes of the
$b$-axis lattice parameter. These results are complemented by
measurements of the hydrostatic pressure dependence of $J_2$ by
means of thermal expansion and magnetic susceptibility measurements
performed both at ambient and finite hydrostatic pressure. We
propose that a structural peculiarity of this compound, in which
Cu$_2$O$_6$ dimer units are incorporated in an unusually stretched manner, is
responsible for the anomalously large magneto-elastic coupling.

\end{abstract}


\maketitle


\section{INTRODUCTION}
Low-dimensional quantum-spin systems have attracted continuous
attention due to the wealth of unusual magnetic properties that
result from the interplay of low dimensionalty, competing
interactions and strong quantum fluctuations. Among these systems,
the diamond chain has been of particular interest, where triangular
arrangements of spin $S$ = 1/2 entities with exchange coupling
constants $J_1$, $J_2$ and $J_3$, are connected to form chains.
\cite{re1, re2, re3, re4} In recent years, great interest has
surrounded the discovery of azurite, Cu$_3$(CO$_3$)$_2$(OH)$_2$,
\cite{structure} as a model system of a Cu$^{2+}$($S$ = 1/2)-based
distorted diamond chain with \textit{$J_1$ $\neq$ $J_2$ $\neq$
$J_3$}.\cite{re5} The observation of a plateau at 1/3 of the
saturation magnetization \cite{re5} is consistent with a description
of azurite in terms of an alternating dimer-monomer model
\cite{re31, re2}. Two characteristic temperatures (energies) have
been derived from peaks in the magnetic susceptibility $\chi$($T$).
\cite{re31, re5} Whereas the peak at $T^{\chi}_{1} \simeq$ 25\,K,
has been assigned to the dominant intra-dimer coupling $J_{2}$, the
one at $T^{\chi}_{2} \simeq$ 5\,K has been linked to a
monomer-monomer coupling along the chain $b$ axis.\cite{re5} There
have been conflicting results, however, as for the appropriate
microscopic description of the relevant magnetic couplings of
azurite.\cite{re6, re7, re8, re9, re10} Very recently, Jeschke
\emph{et al.} \cite{re11} succeeded in deriving an effective
microscopic model capable of providing a consistent picture of most
available experimental data for not too low temperatures, i.e.,
distinctly above the transition into long-range antiferromagnetic
order at $T_N$ = 1.86\,K \cite{re32}. According to this work, the
exchange couplings $J_1$, $J_2$ and $J_3$ are all antiferromagnetic,
thus placing azurite in the highly frustrated parameter regime of
the diamond chain. Within the "refined model" proposed there,
$J_{2}$/$k_{B}$ = 33\,K and an effective monomer-monomer coupling
$J_{m}$/$k_{B}$ = 4.6\,K were found.

Another intriguing property of azurite, not accounted for so far from theory, refers to the strong
magneto-elastic couplings in this compound. These couplings manifest
themselves, e.g., in a pronounced structural distortion accompanying
the magnetic transition at $T_{N}$, as revealed by thermal expansion
\cite{re27, re12} and neutron scattering experiments \cite{re12,
re30}. Here we present a study of these magneto-elastic couplings of
azurite by means of temperature-dependent measurements of the
elastic constant and uniaxial thermal expansion coefficients. These
data are supplemented by thermal expansion and susceptibility
measurements under hydrostatic pressure conditions. The salient
results of our study is the observation of an extraordinarily large
magneto-elastic coupling constant of the intra-dimer coupling
$J_2$ with respect to intra-chain deformations. This coupling
manifests itself in pronounced anomalies in the elastic constant and
uniaxial thermal expansion coefficients, the latter are
characterized by a negative Poisson effect. We propose that the anomalous
magneto-elastic behavior of azurite is a consequence of the material's structural
peculiarities, in particular, the presence of unusually stretched Cu$_2$O$_6$ dimer
units.

\section{EXPERIMENTAL DETAILS}
The single crystals (samples \#1 - \#4) used for the measurements
described in this paper were cut from a large high-quality single
crystal which was also studied by neutron scattering and muon spin
resonance ($\mu$SR) \cite{re7, re12}. For the ultrasonic experiments
two parallel surfaces normal to the [010] direction were prepared
and two piezoelectric polymer-foil transducers were glued to these
surfaces. Longitudinal sound waves for frequencies around 75\,MHz
were propagated along the [010] direction to access the acoustic
$c_{22}$ mode. By using a phase-sensitive detection technique
\cite{re15} the relative change of the sound velocity and the sound
attenuation were simultaneously measured as the function of
temperature for 0.08\,K $\leq T \leq$ 310\,K. A top-loading dilution
refrigerator was used for measurements at $T \leq$ 1.5\,K, whereas a
$^{4}$He bath cryostat, equipped with a variable temperature insert,
was employed for accessing temperatures $T \geq$ 1.5\,K. The elastic
constant $c_{ij}$ is obtained from the sound velocity $v_{ij}$ by
$c_{ij}$ = $\rho v_{ij}^2$ where $\rho$ is the mass density. For
measurements of the uniaxial thermal expansion coefficients,
$\alpha_i(T)$ = $l_i^{-1}$($\partial l_i$/$\partial T$), where
$l_i(T)$ is the sample length along the $i$ axis, two different
dilatometers were used. Experiments under ambient pressure along the
$a'$, $b$ and $c^*$ axes, where $a'$  and $c^*$ are perpendicular to
the ($\overline{1}$02) and (102) crystallographic planes,
respectively, were carried out by means of an ultrahigh-resolution
capacitive dilatometer, built after ref.\,\onlinecite{re16}, with a
resolution of $\Delta l/l \geq$ 10$^{-10}$. In addition,
measurements along the $b$ axis were performed by using a different
dilatometer \cite{re22}, with a slightly reduced sensitivity of
$\Delta l/l \geq$ 5$\cdot$10$^{-10}$, enabling measurements to be
performed under Helium-gas pressure. The magnetic susceptibility at
ambient pressure and at various finite pressure values was measured
with a SQUID magnetometer (Quantum Design MPMS). For the
measurements under pressure, a CuBe piston cylinder clamped cell was
used with Daphne oil 7373 as a pressure-transmitting medium. At low
temperature, the pressure inside the pressure cell was determined by
measuring the superconducting transition temperature of a small
piece of Indium.

\section{RESULTS AND DISCUSSION}
\subsection{Elastic anomalies and pressure/strain dependence of the relevant magnetic energy scales}

\begin{figure}
\includegraphics[width=1.0\columnwidth]{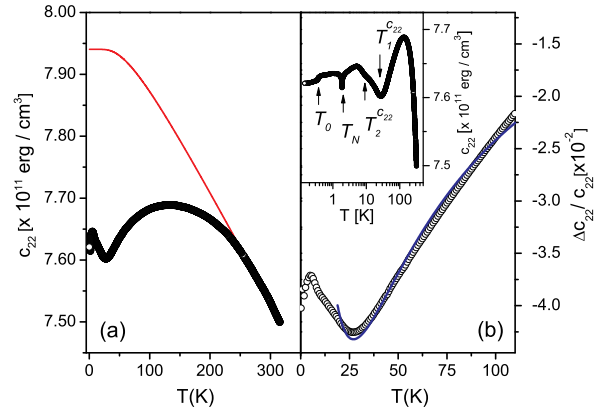}
\caption{\label{hinh1}(Color online). (a) Temperature dependence of
the longitudinal elastic $c_{22}(T)$ mode (open circles) of single crystalline azurite (sample \#1).
The red solid line represents the non-magnetic elastic background
$c_{bg}$($T$) derived from fitting eq.\,\ref{eq1}  to the experimental data
for $T\geq$ 250\,K, see text for details. (b) Normalized magnetic
contribution to the elastic $c_{22}$ mode $\Delta$$c_{22}(T)$  =
[$c_{22}(T) - c_{bg}$($T$)]/$c_{bg}$($T=0$) as a function of
temperature for $T \leq$ 125\,K. The solid blue line represents a
fit to the experimental data (open symbols) based on eq.\,\ref{eq2}.
Inset to (b): The longitudinal mode $c_{22}(T)$ on a logarithmic
temperature scale. The arrows mark the positions of the elastic anomalies in $c_{22}$
of azurite which are connected to the various energy scales ($T^{c_{22}}_1$, $T^{c_{22}}_2$), the phase transition into
long-range antiferromagnetic order ($T_N$), and a transition of unknown, most likely magnetic origin ($T_0$). }
\end{figure}

Figure \ref{hinh1}(a) shows the experimental results (open symbols)
of the longitudinal elastic constant $c_{22}(T)$ of azurite (sample
\#1) over the whole temperature range investigated. Upon cooling,
the $c_{22}$ mode initially increases (hardening) as expected for
materials where anharmonic phonon interactions dominate. Upon
further cooling, however, a pronounced softening becomes visible below
about 250\,K which is accompanied by various anomalies at
lower temperatures $T \leq$ 30\,K. These anomalies can be discerned
particularly clearly in the inset of Fig.\,\ref{hinh1}(b), where the
low-temperature data are shown on a logarithmic temperature scale.
Most prominent is a distinct minimum at a temperature around 27\,K.
Moreover, the data disclose a small dip slightly below 10\,K. The
position of this feature is difficult to estimate due to its
smallness and the strong variation of $c_{22}$ with temperature
caused by the nearby anomalies. We assign these features (labeled
$T^{c_{22}}_1$ and $T^{c_{22}}_2$ in the inset to
Fig.\,\ref{hinh1}(b)) to the characteristic temperatures $T_1$ and
$T_2$ of azurite, as revealed by susceptibility
measurements\cite{re5,re31}. In addition, the elastic data highlight
a sharp minimum around 1.9\,K, reflecting the transition into long-range
antiferromagnetic ordering at $T_{N}$ = 1.88\,K, and a step-like
softening of comparable size to that at $T_{N}$ around $T_{0}$ =
0.37\,K. The latter feature is likely to be of magnetic origin as
well \cite{re23}.

For a quantitative analysis of the $c_{22}$ data, we determined the
magnetic contribution by subtracting the non-magnetic (normal) elastic background
$c_{bg}$. For $c_{bg}$ a phenomenological
expression

\begin{equation}
\emph{c}_{bg}(T) = \emph{c}_{bg}^0 - \frac{\emph{s}}{\emph{e}^{\frac{t}{T}}-1}
\label{eq1}
\end{equation}

was used \cite{re17} which is fitted to the experimental data at
temperatures high enough so that magnetic interactions can be
neglected. Here $\emph{c}_{bg}^0$ is the value of the elastic
constant at $T$ = 0, and $s$ and $t$ are constants. The quantity $t$
is usually set to $\Theta_D$/2, where $\Theta_D$ is the Debye
temperature (see ref.\,\onlinecite{re17} for details). By choosing
$t$ = 175\,K, corresponding to $\Theta_D$ = 350\,K as derived from
specific heat \cite{re18}, and by using ${c}_{bg}^0$ and $s$ as free
parameters, eq.\,\ref{eq1} was fitted to the data for temperatures
250\,K $\leq T \leq $ 310\,K. The elastic background obtained by
this fitting procedure is displayed as the solid red line in
Fig.\,\ref{hinh1}(a).
\begin{figure}
\includegraphics[width=1.0\columnwidth]{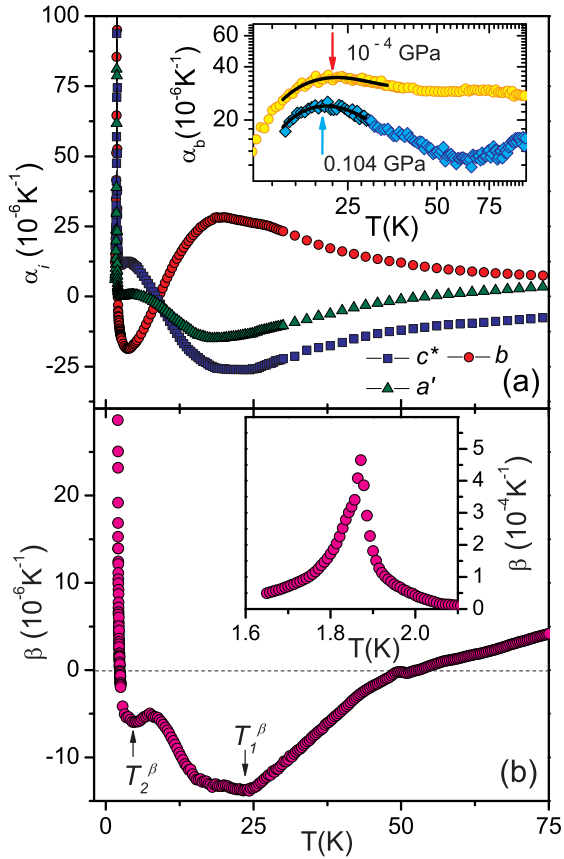}
\caption{\label{hinh2}(Color online). (a) Uniaxial coefficients of thermal expansion of
single crystalline azurite (sample \#3) along the chain $b$ axis and two orthogonal axes perpendicular to $b$ labeled
$a'$, $c^*$. Inset: Temperature dependence of the $b$-axis
expansivity measured on crystal \#4 for 20\,K $\leq T \leq$ 100\,K
at ambient pressure (10$^{-4}$\,GPa) (yellow circles) and at a Helium-gas
pressure of 0.104\,GPa (blue diamonds) in a double-logarithmic
representation, cf. ref. \onlinecite{re22}. The arrows mark the position of
the local maximum in $\alpha_b$ at $T_1$ derived from the
fourth-order polynomial (black solid line) fitted to the
experimental data. (b) Volume expansivity, $\beta$ = $\alpha_{a'}$ +
$\alpha_b$ + $\alpha_{c^*}$ determined from the data in (a). Arrows
labeled $T^{\beta}_1$ and $T^{\beta}_2$ mark the position of
anomalies (minima) in $\beta$. Inset: Details of the phase transition anomaly
in $\beta$ at $T_N$.}
\end{figure}
Fig.\,\ref{hinh1}(b) shows the temperature dependence of the
magnetic contribution to the elastic constant $c_{22}(T)$ for $T
\leq$ 125\,K obtained by subtracting the elastic background from the
experimental data. The so-derived magnetic contribution $\Delta c_{22}$ reveals a
large softening of more than four percent on cooling down to $T^{c_{22}}_1$.
This effect indicates an extraordinarily strong magneto-elastic
coupling which is likely of exchange-striction type for the
following reason. For longitudinal modes, such as the $c_{22}$ mode
investigated here, a two-ion magneto-elastic coupling arises from
the modulation of the distance or bond angles between the magnetic ions which changes the interaction. Furthermore
the single-ion magneto-elastic coupling for Cu$^{2+}$ is small
because of the vanishing quadrupole matrix elements \cite{re13}. In
order to quantitatively evaluate the corresponding magneto-elastic
coupling constant, we introduce in eq.\,\ref{eq2} a generalized
strain susceptibility $\chi_{str}$. This model accounts, within a
random-phase approximation (RPA) \cite{re13}, for the temperature
dependence of the elastic constants of coupled dimers characterized
by an intra-dimer coupling constant corresponding to the dimers'
singlet-triplet excitation gap $\Delta$. This results in a
temperature dependence for the elastic constant \cite{re13} of:

\begin{equation}
\emph{c}_{22}(T) = \frac{\partial^2 {F}}{\partial ({\varepsilon_{22})^2}} = \emph{c}_{bg}(T)-NG^2\chi_{str}(T)
\label{eq2}
\end{equation}

where $F$ is the free energy, $\varepsilon_{22}$ is the strain along the [010] direction, and $G$ =
$\partial\Delta$/$\partial\varepsilon_b$ is the variation of the
singlet-triplet energy gap $\Delta$ of the dimers upon applying a
$b$-axis strain, i.e., $G$ = $\partial J_2$/$\partial\varepsilon_b$
for the case of azurite. $N$ is the density of dimers and

\begin{equation}
\chi_{str}(T) = \frac{\chi_s(T)}{1-K\chi_s(T)},
\label{eq3}
\end{equation}

is the generalized strain susceptibility. Here $\chi_s$($T$)\,=\,
3e$^{-\Delta/k_BT}$/($k_B$$T$$Z^2$) denotes the strain susceptibility of
a single dimer with $Z$ \,=\,1+3e$^{-\Delta/k_BT}$ the partition
function and $K$ the strength of the effective magneto-elastic
dimer-dimer interaction.

The solid blue line in Fig.\,\ref{hinh1}(b) shows a fit to the
experimental data using eq.\,\ref{eq2} with a dimer density $N$ =
0.9918$\cdot$10$^{22}$ cm$^{-3}$, corresponding to 2 dimers per unit
cell, and $c_{22}^0(T = 0)$ = 7.94$\cdot$10$^{11}$ erg/cm$^3$, the
latter is taken from the fitting procedure of the elastic
background. From a nonlinear least-squares fit the parameters
$\Delta$ = $J_2$, $K$ and $|G|$ can be derived. The fit was
constrained to the temperature range 20\,K $\leq T \leq$ 125\,K
where the intra-dimer coupling represents the dominant magnetic
interaction in the system. The so-derived curve provides a good
overall description of the data in the selected temperature window,
especially it reproduces well the observed minimum at $T^{c_{22}}_1
\sim$ 27\,K. The deviations of the fit from the data become
significant at temperatures $T$ $\leq$ 22 K (not shown in
Fig.\,\ref{hinh1}(b)). This is attributed to the simultaneous action
of additional magnetic couplings at lower temperatures, especially
those related to $T_2$. From the fit we obtain $J_2$ = (57 $\pm$
5)\,K and a coupling constant $|G| \sim$ (3650 $\pm$ 150)\,K. The
so-derived coupling constant $J_2$ is somewhat larger than the one
found in the DFT calculations \cite{re11}. Most remarkable, however,
is the exceptionally large value obtained for the coupling constant
$|G|$, reflecting an extraordinarily large strain dependence of the
dominant energy scale in azurite. This value, corresponding to a
softening of approximately 4$\%$, exceeds the coupling constants
typically revealed for other low-dimensional quantum spin systems
\cite{Poirier, BW2004} by one to two orders of magnitude. It is even
about four times bigger than the very large coupling constant found
for the longitudinal mode in the coupled-dimer system
SrCu$_2$(BO$_3$)$_2$ \cite{re21}. For the parameter $K$ in
eq.\,\ref{eq3} the fit yields $-$(200 $\pm$ 10)\,K, indicating that
the effective magneto-elastic dimer-dimer
interaction is antiferrodistortive in azurite. \\

In order to obtain supplementary information about the magneto-elastic couplings corresponding to the
various energy scales, we performed measurements of the coefficient of thermal expansion and the
magnetic susceptibility both at ambient- and under hydrostatic-pressure conditions.

Figure \ref{hinh2}(a) shows the results of the temperature dependence
of thermal expansion coefficients $\alpha_i$($T$) ($i$ = $a'$, $b$,
$c^*$) for temperatures 1.6\,K $\leq T \leq$ 75\,K measured at
ambient pressure. Similar to the elastic constant, three distinct
anomalies are observed in this temperature range along all three
crystallographic axes. Upon cooling, the $b$-axis data show a broad
positive maximum around 20\,K whereas along the two other axes ($a'$
and $c^*$) a negative minimum appears. Upon further cooling to lower
temperatures, a minimum shows up in the $b$-axis data, which
contrasts with maxima around 3.5\,K in the data along the $a'$- and
$c^*$-axis. Despite having opposite signs, the anomalies in
$\alpha_b(T)$ and those in $\alpha_{a'}(T)$ and $\alpha_{c^*}(T)$ do
not cancel each other out in the volume expansion coefficient
$\beta$($T$) =  $\alpha_{a'}(T)$ + $\alpha_b(T)$ +
$\alpha_{c^*}(T)$, shown in the main panel of Fig.\,\ref{hinh2}(b).
The volume expansivity exhibits a pronounced negative contribution,
giving rise to a change of sign around 50\,K, and a broad minimum
around 25\,K followed by a second minimum at $\simeq$ 5\,K. The strong upturn in $\beta$($T$) and $\alpha_i$ at
lower temperatures is due to the antiferromagnetic phase transition
at $T_N$. This is shown more clearly in the inset of
Fig.\,\ref{hinh2}(b) where $\beta$($T$) is displayed on expanded
scales around $T_N$. An extraordinarily large $\lambda$-type anomaly, lacking any hysteresis upon cooling and warming, is observed
which demonstrates the second-order character of the phase
transition at $T_N$. From the coincidence of the anomalies in
$\beta$($T$) (and $\alpha_i$($T$)) with those revealed in the magnetic
susceptibility \cite{re5} and elastic constant (Fig.\,\ref{hinh1}), we conclude that these anomalies reflect the
characteristic temperatures $T_1$ and $T_2$ which are related to the
energy scales $J_2$ and $J_{m}$, respectively. Thus from the
evolution of these anomalies under pressure we may determine the
pressure dependencies of these energy scales. In the inset of
Fig.\,\ref{hinh2}(a) we compare, on a logarithmic temperature scale,
the temperature dependence of $\alpha_b$ for a small pressure of at
10$^{-4}$\,GPa with that of 0.104\,GPa. These data, which have been taken on
sample \#4 by employing a different dilatometer, especially
designed for measurements under Helium-gas pressure \cite{re22},
disclose two remarkable features. First, we find a strong change in
the temperature dependence of $\alpha_b$($T$), accompanied by a
considerable suppression in its absolute value over the whole
temperature range investigated, i.e., for $T \leq$ 100\,K. Note that
due to the solidification of the pressure medium, the measurements
at 0.104\,GPa were limited to $T >$ 14.5\,K. Since changes of the
lattice expansivity under a pressure of 0.104\,GPa for a material with
a normal bulk modulus are expected to be less than 1\%, see
ref.\,\onlinecite{re22} and the discussion therein, we attribute these
effects to the influence of pressure on the material's magnetic
properties and their coupling to the lattice degrees of freedom. The
presence of significant magnetic contributions at elevated
temperatures $T \gg J_2/k_B$ which strongly couple to the lattice is
consistent with the acoustic behavior revealed for the $c_{22}$($T$)
elastic mode, yielding an onset temperature for the pronounced
softening as high as 250\,K, cf., Fig.\,\ref{hinh1}(a). Second, a
thorough inspection of the data around the maximum reveals a shift
of the position of the maximum to lower temperatures on increasing
the pressure from $p$ = 0.1 to 0.104\,GPa. This can be quantified more
precisely by fitting both data sets in the immediate surrounding of the maximum by fourth-order polynomials
\cite{re22}, depicted as solid lines in Fig.\,\ref{hinh2}(a). By
identifying the position of the maximum with $T_1$, we find a
pressure dependence $\partial T_1$/$\partial p$ = $-$(0.08 $\pm$ 0.03)\,K/GPa.

\begin{figure}
\includegraphics[width=1.0\columnwidth]{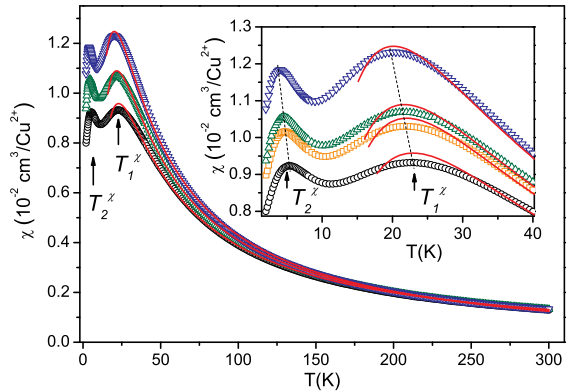}
\caption{\label{hinh3}(Color online). Temperature dependence of
magnetic susceptibility of single crystalline azurite (sample \#2)
measured at various hydrostatic pressure values (from bottom to top ambient pressure, 0.36, 0.62\,GPa) in a magnetic field
of $B$ = 2\,T applied parallel to the $b$-axis. Red solid lines
correspond to fits to the data based on a model for interacting
dimers given in eq.\,\ref{eq4}. Inset: Blow-up of the data (symbols)
and model curves (red lines) in the vicinity of the characteristic
temperature $T_1^\chi$. Besides the data at  ambient pressure (black circles), $p$ = 0.360\,GPa (green triangles) and 0.62\,GPa (purple triangles), also data at 0.26\,GPa (orange squares) are shown. Arrows labeled  $T^{\chi}_1$
and $T^{\chi}_2$ mark the position of the maxima in the various data
sets. Broken lines indicate the shift in the maximum position with
pressure.}
\end{figure}

Figure \ref{hinh3} displays the data of the magnetic susceptibility
$\chi$($T$, $p$ = const) as a function of temperature for 2\,K $\leq
T \leq$ 300\,K at varying pressure values from ambient pressure up to
0.62\,GPa. The ambient-pressure data are consistent with those
reported by Kikuchi \textit{et al.} \cite{re5}, yielding an increase
in $\chi$ with decreasing temperature and two broadened maxima at
$T^{\chi}_1$ and $T^{\chi}_2$. Upon increasing the pressure we
observe a progressive increase of the low-temperature
susceptibility, cf.\,Fig.\,\ref{hinh3}. In addition, a closer look
at the low-temperature data in the inset of Fig.\,\ref{hinh3}
discloses a shift of the position of both maxima to lower
temperatures albeit at different rates.

For a quantitative analysis of the susceptibility data for not too
low temperatures around $T_1$ and up to 300\,K, we again use an
RPA-molecular field expression for coupled dimers, in analogy to the
procedure applied for analyzing the elastic constant $c_{22}(T)$
data:

\begin{equation}
\chi_{m}(T) = \frac{\chi_0 (T)}{1 - \widetilde{K}\chi_0(T)}.
\label{eq4}
\end{equation}

Here $\chi_0$($T$) = 2e$^{-\Delta/k_BT}/(k_BTZ)$ is the magnetic
susceptibility of an isolated dimer, $Z$ the partition function and
$\widetilde{K}$ an average magnetic inter-dimer coupling. A fit
to the experimental data at ambient pressure was performed for 15\,K $\leq T \leq$ 300\,K by using eq.\,\ref{eq4} and a Curie contribution, according to
the amount of Cu-monomer spins, with $\Delta$ = $J_2$ and
$\widetilde{K}$ as free parameters. This fit provides a very good description of the data
including the height and the position of the maximum at
$T^{\chi}_1$, cf.\,\ the solid red line running through the ambient-
pressure data points in Fig.\,\ref{hinh3}. From the fit we obtain
$J_2$/$k_B$ = (40.5 $\pm$ 0.7)\,K, consistent with the value
suggested by DFT calculations for the "full model" discussed there \cite{re11}, and
$\widetilde{K}$ = (4.1 $\pm$ 0.8)\,K. Note that this value of
$\widetilde{K}$ indicates that the average magnetic dimer-dimer
interaction is small and ferromagnetic. The RPA-molecular field
description remains good also for the data taken at finite pressure. For these fits the same Curie susceptibility as used for the ambient-pressure data was used to account for the magnetic contributions of the monomers.
The evolution of the parameters $J_2$ and $\widetilde{K}$ with
pressure, derived from fitting eq.\,\ref{eq4} to the finite-pressure data, is shown in Fig.\,4. We
find a suppression of $J_2$ with increasing pressure which can be
approximated by $\partial J_2$/$\partial p$ $\approx$ $-$(0.077 $\pm$
0.007)\,K/GPa. For the average magnetic inter-dimer interaction
we find an approximate linear increase under pressure with a
larger rate of $\partial \widetilde{K}$/$\partial p \approx$ (0.216
$\pm$ 0.002)\,K/GPa. This pronounced increase of $\partial
\widetilde{K}$/$\partial p$ is responsible for the growth of the low-temperature susceptibility with pressure.

 As will be discussed in more detail below (Sec.\,C), the dominant ferromagnetic character of the dimer-dimer
interaction can be assigned to a particular Cu-O-Cu exchange path connecting
dimers of adjacent chains along the $a$-direction. \\

\begin{figure}
\includegraphics[width=1.0\columnwidth]{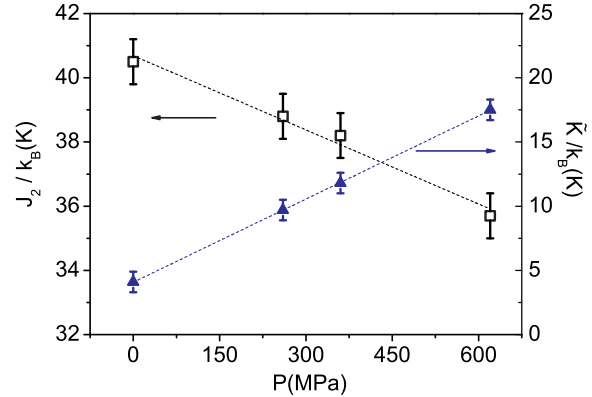}
\caption{\label{hinh4}(Color online). Evolution of the fit
parameters, the intra-dimer coupling constant $J_2$ (open squares, left scale) and the average
magnetic dimer-dimer interaction energy $\widetilde{K}$ (full triangles, right scale), with
pressure as derived from fits of eq.\,\ref{eq2} to the data in
Fig.\,3.}
\end{figure}

\subsection{Magneto-elastic coupling at $T_N$}

\begin{figure}
\includegraphics[width=1.0\columnwidth]{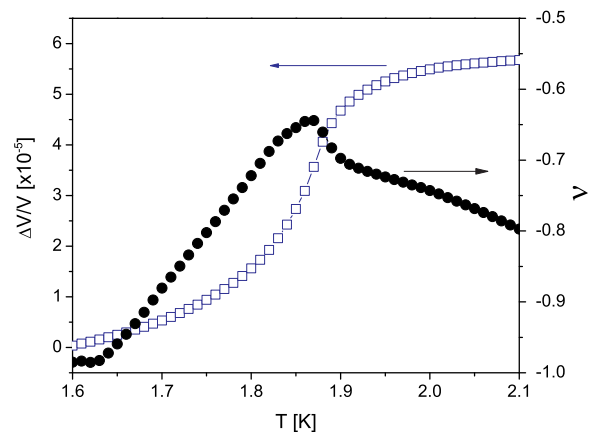}
\caption{\label{hinh4}(Color online). Poisson ratio (black full
circles, right scale) together with the relative volume change around
$T_N$ (blue open squares, left scale).}
\end{figure}

As shown in the previous section, the phase transition into the
long-range antiferromagnetic order in azurite at $T_N$  = 1.88\,K is
accompanied by a pronounced anomaly in the elastic $c_{22}$ mode,
corresponding to a softening of about 0.1$\%$ (cf.\,inset to
Fig.\,\ref{hinh1}(b)) which exceeds by far the features usually
revealed at a magnetic transition in low-dimensional spin systems
where often only a kink-like anomaly is observed, see
refs.\,\onlinecite{Poirier}, \onlinecite{re26}, for typical
examples, and ref.\,\onlinecite{re13} for a review. This goes along
with an extraordinarily large anomaly in the uniaxial coefficients
of thermal expansion $\alpha_i$, see Fig. \ref{hinh2}(b). As a
result of the strong magneto-elastic coupling, there is a large
$\lambda$-type anomaly in the volume expansion coefficient $\Delta
\beta$, (see the inset of Fig.\,\ref{hinh2}(b)), corresponding to a
relative reduction of the volume upon cooling from 2 to 1.6\,K of
$\Delta V/V$ = $-$5.7 $\cdot$
10$^{-5}$ as shown in Fig.\,5.\\

According to the Ehrenfest relation, the
discontinuities at this second-order phase transition in $\beta$,
$\Delta \beta$, and that in the specific heat, $\Delta C_p$, can be
used to determine the pressure dependence of the N\'{e}el
temperature in the limit of vanishing pressure

\begin{equation}
\left(\frac{\partial T_N}{\partial p}\right)_{p\rightarrow 0} = V_{mol}\,T_{N}\,\frac{\Delta \beta}{\Delta C_p}.
\label{eq6}
\end{equation}

By using $\Delta \beta$ = (550 $\pm$ 30) $\times$ 10$^{-6}$ K$^{-1}$
and $\Delta C_p$ = (6.18 $\pm$ 0.4) $\times$ J mol$^{-1}$K$^{-1}$
taken from ref.\,\onlinecite{re7}, we find $\left(\frac{\partial
T_N}{\partial p}\right)_{p\rightarrow 0}$ = (0.15 $\pm$
0.02)\,K/GPa. This extraordinarily large pressure dependence,
exceeded only by some exceptional cases, such as the coupled
antiferromagnetic/structural transition in Co-substituted
CaFe$_2$As$_2$ \cite{Gati2012}, highlights the strong
magneto-elastic coupling and the unusual elastic properties of azurite.\\

The large $\Delta \beta$ at $T_N$, tantamount to a large pressure dependence of $T_N$ in eq.\,\ref{eq6}, is partly due to the fact
that in azurite at $T_N$ the discontinuities in the uniaxial expansion coefficients,
$\Delta \alpha_i$, all have the same (positive) sign. Hence, this phase transition is characterized
by an anomalous Poisson effect.\\

In general for an isotropic material, the Poisson ratio, which measures the material's cross section under tension, is
defined as:

\begin{equation}
\nu = - \frac{\epsilon_y}{\epsilon_x },
\label{eq5}
\end{equation}

where $\epsilon_x$ is the strain in stretching direction and
$\epsilon_y$ perpendicular to it. In most materials $\nu$ is positive,
which reflects the fact that an expansion along one axis is usually
accompanied by a compression in the perpendicular direction, to keep the
overall volume change small. In many materials $\nu$ values are found in the range
0.2 $\leqslant$ 0.5 \cite{auxetic}. The latter value corresponds to the situation that the material keeps its volume under tension.
A positive $\nu < 0.5$ means that the material becomes thinner when it is stretched, the behavior encountered
for most materials. In contrast, materials with a negative Poisson ratio become thicker when they are stretched.
Those compounds, called auxetic materials, are of interest due to potential technical applications \cite{auxetic}.

In Fig.\,5 we show the Poisson ratio $\nu$ of azurite for temperatures around $T_N$.
For the present anisotropic case, $\nu $ has been determined by using the relative length changes ($\Delta l$/$l$)$_b$
(corresponding to the integral of $\alpha_b$($T$) with respect to temperature) for the
strain $\epsilon_x$ along the stretching $b$ direction and $\overline{\epsilon_y}$ = ($\epsilon_{a'}$ + $\epsilon_{c^{*}}$)/2 = (($\Delta l$/$l$)$_{a'}$ + ($\Delta l$/$l$)$_{c^{*}})/2$ perpendicular
to it. We stress that $\nu$ of azurite reaches a normal value of 0.23 around room
temperature (not shown). However, as indicated in Fig.\,5, $\nu$ exhibits a large negative
value of $-$0.8 for temperatures slightly above $T_N$. This value increases to $-$0.65
upon cooling to $T_N$ below which it further decreases, reaching almost $-$1 at 1.6 K.
Note that this value is close to the stability limit of any elastic linear, isotropic material where the
requirement of positive Young's shear and bulk moduli dictates $\nu$ $>$ $-$1.\\

\subsection{Relationship between structural and magnetic properties}

\begin{figure}
\includegraphics[width=1.0\columnwidth]{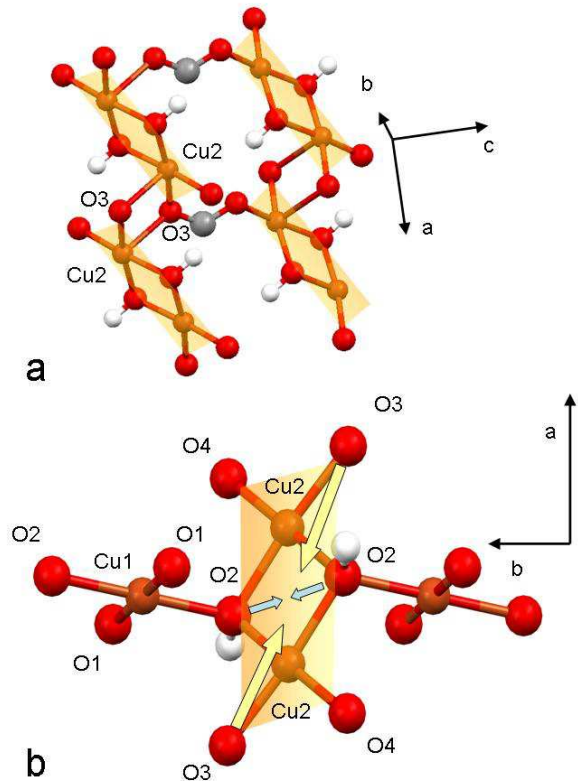}
\caption{\label{hinh4}(Color online). Sections of the structure of azurite. a) The dominant inter-dimer interaction, connecting dimers (marked by yellow planes) in adjacent chains along the $a$-direction, is mediated
via the Cu2-O3-Cu2 exchange path. b) Arrangement of relevant structural units of
azurite including two Cu(II) monomers (CuO$_4$ containing Cu1) and one dimer (Cu$_2$O$_6$ containing Cu2) forming chains
along the $b$-axis viewed perpendicular to the $ab$ plane.
The labels correspond to the room temperature structure reported in ref.\,\onlinecite{re30}.}
\end{figure}

\begin{table*}[htb]
\begin{ruledtabular}
\begin{tabular}{cccc}
{\bf atom 1}-{\bf atom 2}&{\bf distance} {\bf [{\AA}]}&{\bf distance} {\bf [{\AA}]}&{\bf $d_{300K} - d_{5K}$} {\bf [{\AA}]}\\
{\bf }&{\bf at 300\,K} {($d_{300K}$)}&{\bf at 5\,K} {($d_{5K}$)}&{\bf }\\
\hline\hline
O3-Cu2&1.93850&1.93320&0.00530\\
Cu2-O4&1.93990&1.93910&0.00080\\
O2-Cu2&1.99470&1.98820&0.00650\\
Cu2-O2&1.96750&1.96420&0.00330\\
Cu2-Cu2&2.98510&2.99220&-0.00710\\
{\bf O2}-{\bf O2}&{\bf 2.60560}&{\bf 2.58320}&{\bf 0.02240}\\
{\bf O3}-{\bf O3}&{\bf 6.20180}&{\bf 6.19130}&{\bf 0.01050}\\
\end{tabular}
\caption{Structural data of the Cu$_2$O$_6$ dimer units of azurite
at 300\,K and 5\,K taken from refs. \cite{Belokoneva01, re30}. The first column denotes the atoms involved with labels according to Fig.\,6. The second (third) column gives the distance $d$ between these atoms at 300\,K (5\,K). The fourth column gives the difference in these distances upon cooling from 300\,K to 5\,K. The O2-O2 and O3-O3 distances, showing the strongest changes upon cooling and which are involved in the auxetic behavior, are printed in bold.}\label{tab1}
\end{ruledtabular}
\end{table*}
We start the discussion by considering the dimer-dimer interaction $\widetilde{K}$ revealed from the analysis of the susceptibility measurements under variable pressure.
The dominant ferromagnetic character of this interaction is assigned to the Cu2-O3-Cu2 exchange path (cf. Fig. 6a) connecting dimers of adjacent chains along the $a$-axis.
The corresponding Cu2-O3-Cu2 bond angle amounts to 91.57\,$^{\circ}$ at 5\,K, consistent
with a weak ferromagnetic interaction as revealed for
hydroxo-bridged Cu(II) complexes \cite{Crawford76}. Note that also
the DFT calculations for the "full model" of azurite exhibit a small
ferromagnetic exchange \cite{re11}. It is likely that under hydrostatic pressure the structure will deform in a way such that this angle decreases with increasing pressure. This is consistent with an increase of the ferromagnetic inter-dimer coupling $\widetilde{K}$ derived from susceptibility measurements under pressure, cf.\,Fig.\,4. \\

In the following we will argue that the auxetic behavior at $T_N$
and the huge magneto-elastic coupling of azurite is likely due to
peculiarities of the molecular arrangement in this compound, in
particular that of the Cu$_2$O$_6$ dimer units, cf.\,Fig.\,6. As
shown in Fig.\,6b, CuO$_4$ monomer units (containing Cu1) are
connected via O2 ions to Cu$_2$O$_6$ dimers (containing Cu2 and
marked by yellow planes in Fig.\,6) to form chains along the $b$
axis. According to structural data at room temperature
\cite{Belokoneva01} and 5\,K \cite{re30}, cf.\,Table 1, the
structural parameters of the monomers, involving O1-Cu1 and O2-Cu1
bonds, change very little upon cooling and are very close to
those values (1.90 - 1.93\,${\AA}$) typically found in isolated
Cu(II) complexes \cite{Crawford76}. This indicates a stable and
rigid configuration of the CuO$_4$ monomer units. In contrast, the
structural parameters of the Cu$_2$O$_6$ dimers are rather unusual.
In particular the Cu2-O2 bonds, mediating the intra-dimer coupling
$J_2$, are significantly longer (1.9947 ${\AA}$ and 1.9675 ${\AA}$
at 300\,K) (see Table 1) than those found in isolated dimer complexes, reflecting an unusually "stretched" arrangement. It is thus obvious to suspect that this dimer unit represents the flexible part
in the structure which may accommodate itself accordingly when the
surrounding structure is exposed to external stimuli. This is the
case, e.g., when a $b$-axis strain is applied or, alternatively,
upon cooling through $T_1$ which is accompanied by a significant
contraction of the $b$ axis (large positive anomaly in $\alpha_b$,
cf.\,Fig.\,2). In fact, according to structural data (Table\,1), it
is the O2-O2 distance, as indicated by the blue arrows in Fig.\,6b,
across the dimer which yields the strongest reduction upon cooling
from at 300\,K to 5\,K, reflecting the softness of these dimer
units. Note that changes of the Cu2-O2 geometry imply changes of the
dominant magnetic interaction $J_2$. As Fig.\,6b suggests, such a
contraction along the $b$-direction will cause not only a reduction
of the O2-O2 distance. It will also lead to changes of the Cu2-O2
bond length and the Cu2-O2-Cu2 bond angle (97.77$^{\circ}$ at
300\,K) within the dimers. In fact, the 5\,K structure reveals an
Cu2-O2-Cu2 angle of 98.41(7)$^{\circ}$. At the same time it is
conceivable that upon the reorientation of the dimer in response to
a $b$-axis strain, the tensile forces acting on the dimers get
weaker, allowing the dimers to adopt a more natural, compact
configuration, consistent with the observed reduction of the Cu2-O2
bond lengths. This less-strained dimer configuration is also visible
in the intra-dimer O3-O3 distance which is reduced as well (see
Table 1 and the yellow arrows in Fig. 6b). Due to the orientation of the
dimer unit in the $ac$ plane, a reduction of the $b$ axis is connected
with a reduction of the $a$- and $c$-axis
which then explains the anomalous Poisson effect \cite{note} at low temperatures. \\

As for the anomalously large magneto-elastic coupling $|G|$, one
might expect that the stretched arrangement of the dimers may alter
the relationship between the exchange coupling constant and the
dimers' structural parameters $\eta$, such as the inter-atomic
distances or bonding angles, and with it the generalized derivatives
$\partial J$/$\partial \eta$. According to Crawford \textit{et al.}
\cite{Crawford76}, who investigated various stable hydroxo-bridged
dimer complexes, there is a linear correlation between the
intra-dimer coupling and both the Cu-Cu distance as well as the
Cu-O-Cu bond angle. All of these materials exhibit nearly the same
typical Cu-O distances. It is likely that these relations may change
significantly for a strongly stretched configuration, as realized
for the dimer units in azurite. This may result in strongly enhanced
derivatives $\partial J$/$\partial \eta$ such as the large
magneto-elastic coupling constant $\partial J$/$\partial \epsilon_b$
revealed here. A microscopic theory, addressing the relationship between the coupling constants and the dimers' structural parameters, is necessary to confirm this conjecture. Note that the presence of pre-strained structural
units in azurite is consistent with the fact that a hydrothermal
synthesis technique operating at a pressure of about 0.350\,GPa has to be
applied for growing single crystals of this mineral
\cite{Ruszala74}.\\

\section{CONCLUSIONS}
Measurements of the longitudinal elastic constant $c_{22}$ and the
uniaxial thermal expansion coefficients $\alpha_i$ on single
crystalline azurite reveal pronounced anomalies associated with the
intra-dimer coupling constant $J_2$. From a quantitative analysis of
the elastic constant data, an exceptionally large value of the
magneto-elastic coupling $G$ = $\partial J_2$/$\partial \epsilon_b$
of $|G| \sim $ (3650 $\pm$ 150)\,K has been derived. By lacking a
microscopic theory, we tentatively assign this large value,
exceeding corresponding magneto-elastic couplings for other
low-dimensional quantum spin systems by two to three orders of
magnitude, to structural peculiarities of azurite. We propose that
it is the Cu$_2$O$_6$ dimer unit, which is incorporated in the
structure in an unnaturally stretched manner, which is responsible
for the exceptionally large magneto-elastic coupling in this system.

\section{Acknowledgments}
We acknowledge financial support by the Deutsche Forschungsgemeinschaft via the SFB/TR49 and B. L\"{u}thi, R. Valent\'{\i} and H. Jeschke for useful discussions.


\begin{thebibliography}{000}
\bibitem[$\dag$]{presentaddress} Present address: Departamento de F{\'\i}sica, Universidade Estadual Paulista, Rio Claro (SP), Brazil.

\setcounter{NAT@ctr}{000}

\bibitem{re1} K. Takano, K. Kubo, and H. Sakamoto, J. Phys. Condens. Matter, {\bf 8}, 6405 (1996).

\bibitem{re2} K. Okamoto, T. Tonegawa, Y. Takahashi, and M. Kabugari, J. Phys. Condens. Matter {\bf 35}, 5979 (2003).

\bibitem{re3} T. Tonegawa, K. Okamoto, T. Hikihara, Y. Takahashi, and M. Kaburagi, J. Phys. Soc. Jpn. {\bf 69}, Suppl. A, 332 (2000).

\bibitem{re4} T. Tonegawa, K. Okamoto, T. Hikihara, Y. Takahashi, and M. Kaburagi, J. Phys. Chem. Solids {\bf 62}, 125 (2001).

\bibitem{structure} Azurite crystallizes in the monoclinic space group P2$_1$/c, first discovered by Gattow and Zeeman \cite{gattow}
and later confirmed by two independent investigations \cite{zigan,
Belokoneva01}.

\bibitem{gattow} G. Gattow and J. Zemann, Acta Cryst.{\bf 11}, 866 (1958).

\bibitem{zigan} F. Zigan and H. D. Schuster, Z. Kristallogr. , {\bf 135}, 416 (1972).

\bibitem{Belokoneva01} E. L. Belokoneva, Yu. K. Gubina, and J. B. Forsyth, Phys. Cem. Minerals {\bf 28}, 498 (2001).

\bibitem{re5} H. Kikuchi, Y. Fujii, M. Chiba, S. Mitsudo, T. Idehara, T. Tonegawa, K. Okamoto, T. Sakai, T. Kuwai, and H. Ohta, Phys. Rev. Lett. {\bf 94}, 227201 (2005).

\bibitem{re31} E. Frikkee and J. Handel, Physica {\bf 28}, 269 (1962).

\bibitem{re6}   H. J. Mikeska and C. Luckmann, Phys. Rev. B {\bf 77}, 054405 (2008).

\bibitem{re7}   K. C. Rule, A. U. B. Wolter, S. S\"{u}llow, D. A. Tennant, A. Br\"{u}hl, S. K\"{o}hler, B. Wolf, M. Lang, and J. Schreuer, Phys. Rev. Lett. {\bf 100}, 117202 (2008).

\bibitem{re8}   B. Su and G. Su, Phys. Rev. Lett.  {\bf 97}, 089701 (2006).

\bibitem{re9}   Y. C. Li, J. Appl. Phys {\bf 102}, 113907 (2007).


\bibitem{re10} J. Kang, C Lee, R. K. Kremer,  and M-H Whangbo, J. Phys. Condens. Matter {\bf 21}, 392201 (2009).

\bibitem{re11} H. Jeschke, I. Opahle, H. Kandpal, R. Valenti, H. Das, T. Saha-Dasgupta, O. Janson, H. Rosner, A. Br\"{u}hl, B. Wolf, M. Lang, J. Richter, S. Hu, X. Wang, R. Peters, T. Pruschke, and A. Honecker, Phys. Rev. Lett. {\bf 106}, 217201 (2011).

\bibitem{re32} R. D. Spence and R. D. Ewing, Phys. Rev. {\bf 112}, 1544 (1958).

\bibitem{re27} F. Wolff-Fabris, S. Francoual, V. Zapf, M. Jaime, B. Scott, S. Tozer, S. Hannahs, T. Murphy, and A. Lacerda, J. Phys. Conf. Ser. {\bf 150}, 042030 (2009).

\bibitem{re12} M. C. R. Gibson, K. C. Rule, A. U. B. Wolter, J.-U. Hoffmann, O. Prokhnenko, D. A. Tennant, S. Gerischer, M. Kraken, F. J. Litterst, S. S\"{u}llow, J. Schreuer, H. Luetkens, A. Br\"{u}hl, B. Wolf, and M. Lang, Phys. Rev. B {\bf 81}, 140406(R) (2010).

\bibitem{re30} K. C. Rule, M. Reehuis, M. C. R. Gibson, B. Ouladdiaf, M. J. Gutermann, J.-U. Hoffmann, S. Gerischer, D. A. Tennant, S. S\"{u}llow, and M. Lang, Phys. Rev. B {\bf 83}, 104401 (2011).

\bibitem{re15} B. L\"{u}thi, G. Bruls, P. Thalmeier, B.Wolf, D. Finsterbusch, and I. Kouroudis, J. Low Temp. Phys. {\bf 95}, 257 (1994).

\bibitem{re16} R. Pott and R. Schefzyk, J. Phys. E {\bf 16}, 444 (1983).

\bibitem{re22} R. S. Manna, B. Wolf, M. de Souza, and M. Lang, Rev. Sci. Instrum. {\bf 83}, 085111 (2012).

\bibitem{re23} P. T. Cong, B. Wolf, U. Tutsch, K. Removic-Langer, J. Schreuer, S. S\"{u}llow, and M Lang, J. Phys. Conf. Ser. {\bf 200}, 012226 (2010).


\bibitem{re17} Y. P. Varshni, Phys. Rev. B {\bf 2}, 3952 (1970).

\bibitem{re18} H. Kikuchi, Y. Fujii, M. Chiba, S. Mitsudo, T. Idehara, T. Tonegawa, K. Okamoto, T. Sakai, T. Kuwai, K. Kindo, A. Matsuo, W. Higemoto, K. Nishiyama, M. Horvatic, and C. Bertheir, Prog. Theor. Phys. Suppl. {\bf 159}, 1 (2005).

\bibitem{re13} B. L\"{u}thi, \emph{Physical Acoustics in the Solid State} (Springer, Berlin, 2005).

\bibitem{BW2004} B. Wolf, S. Zherlitsyn, B. L\"{u}thi, N. Harrison, U. L\"{o}w, V. Pashchenko, M. Lang, G. Margraf, H. W. Lerner, E. Dahlmann, F. Ritter, W. Assmus, and M. Wagner, Phys. Rev. B {\bf 69}, 092403 (2004).

\bibitem{Poirier} G. Quirion, T. Taylor, M. Poirier, Phys. Rev. B {\bf 72}, 094403 (2005).

\bibitem{re21} B. Wolf, S. Zherlitsyn, S. Schmidt, B. L\"{u}thi, H. Kageyama, and Y. Ueda, Phys. Rev. Lett.  {\bf 86}, 4847 (2001).

\bibitem{re26} A. Sytcheva , O. Chiatti, J. Wosnitza, S. Zherlitsyn , A. A. Zvyagin, R. Coldea, and Z. Tylczynski, J. Low Temp. Phys. {\bf 159}, 109 (2010).

\bibitem{Crawford76} V.H. Crawford, H. W. Richardson, J. R. Wasson, D.J. Hodgson, and W. E. Hatfield, Inorg. Chem. {\bf 15}, 2107 (1976).

\bibitem{Gati2012} E. Gati, S. K\"{o}hler, D. Guterding, B. Wolf, S. Kn\"{o}ner, S. Ran, S.L. Bud'ko, P.C. Canfield, and M. Lang, Phys. Rev. B {\bf 86}, 220511(R) (2012).

\bibitem{auxetic} K. E. Evans, Endeavour, New Series {\bf 15}, 170 (1992).

\bibitem{note} Note that the anomalous length changes around $T_N$ discussed above,
giving rise to a negative Poisson effect at low temperatures, are of the order of
$\Delta l$/$l$ $\leq$ 10$^{-5}$ and therefore beyond the resolution of standard structural investigations.

\bibitem{Ruszala74} F. Ruszala and E. Kostiner, J. Cryst. Growth {\bf 26}, 155 (1974).


\end{thebibliography}
\end{document}